\documentclass[english,aps,floats,onecolumn,showpacs,nofootinbib]{revtex4}
\usepackage{pslatex}
\usepackage[T1]{fontenc}
\usepackage[latin1]{inputenc}
\usepackage{graphicx}
\usepackage{epsfig}

\usepackage{calc}
\usepackage{ifthen}

{
{
{
\newcommand{\bea}{\begin{eqnarray}}
\newcommand{\eea}{\end{eqnarray}}

\newcommand{\nc}{\newcommand}
\nc{\renc}{\renewcommand}
\nc{\eqs}[2]{\mbox{Eqs.~(\ref{#1},\,\ref{#2})}}
\nc{\eq}[1]{\mbox{Eq.~(\ref{#1})}}
\nc{\figs}[2]{\mbox{Figs.~(\ref{#1},\,\ref{#2})}}
\nc{\fig}[1]{\mbox{Fig~.(\ref{#1})}}
\nc{\be}[1]{\begin{equation} \mbox{$\label{#1}$}}
\nc{\ee}{\vspace{0.1cm}\end{equation}}

\newcommand{\bean}{\begin{eqnarray*}}
\newcommand{\eean}{\end{eqnarray*}}

\def\GeV{{\rm \ GeV}}

\def\lae{\;^{<}_{\sim} \;} \def\gae{\; ^{>}_{\sim} \;}

\begin{document}
\title{Sub-Planckian Two-Field Inflation Consistent with the Lyth Bound}
\author{John McDonald}
\email{j.mcdonald@lancaster.ac.uk}
\affiliation{Dept. of Physics, University of 
Lancaster, Lancaster LA1 4YB, UK}

\begin{abstract}

    The BICEP2 observation of a large tensor-to-scalar ratio, $r = 0.20^{+0.07}_{-0.05}$, implies that the inflaton $\phi$ in single-field inflation models must satisfy $\phi \sim 10M_{Pl}$ in order to produce sufficient inflation. This is a problem if interaction terms suppressed by the Planck scale impose a bound $\phi \lae M_{Pl}$.  Here we consider whether it is possible to have successful sub-Planckian inflation in the case of two-field inflation. The trajectory in field space cannot be radial if the effective single-field inflaton is to satisfy the Lyth bound. By considering a complex field $\Phi$, we show that a near circular but aperiodic modulation of a $|\Phi|^{4}$ potential can reproduce the results of $\phi^2$ chaotic inflation for $n_{s}$ and $r$ while satisfying $|\Phi| \lae 0.01 M_{Pl}$ throughout.
More generally, for models based on a $|\Phi|^{4}$ potential, the simplest sub-Planckian models are equivalent to $\phi^{2}$ and $\phi^{4/3}$ chaotic inflation.

\end{abstract}


 \maketitle

\section{Introduction}

    The observation by BICEP2 of gravity waves from inflation \cite{bicep1,bicep2} is problematic for single-field inflation models. The Lyth bound \cite{lythb}, following an earlier observation of Starobinsky \cite{staro} and recently generalized in \cite{antusch}, implies that the change of the inflaton field during inflation must be $\sim 10 M_{Pl}$ in order to generate enough inflation\footnote{$M_{Pl} = 2.4 \times 10^{18} \GeV$.}. This is a serious problem if there are interaction terms in the inflaton potential suppressed by the Planck mass scale, since these will impose an upper bound $\phi \lae M_{Pl}$. It may be possible to suppress such interactions via a symmetry. However, this requires a quite non-trivial symmetry, for example a shift symmetry \cite{shift,lyths,kaloper} which is slightly broken to allow an inflaton potential \cite{shift}. If Planck-suppressed interactions are not suppressed by a symmetry then the only alternative is to consider multi-field inflation.  One possibility is N-flation\footnote{An alternative multi-field approach which may resolve the problem, M-flation, is described in \cite{matrix}.} \cite{nflation}. In this case, to achieve $\phi_{i} < M_{Pl}$ 
for the $i = 1,...,N$ scalar fields in chaotic $\phi^2$
N-flation, more than 200 scalar fields are necessary \cite{nflation}.   

    Should it be that no symmetry exists to suppress Planck corrections to the inflaton potential and there is not a very large number of scalar fields, then we need to consider an alternative model of inflation which has sub-Planckian values of the scalar fields throughout.

     Here we consider whether this is possible in the case of two-field inflation. In general, this will require an unconventional potential. We will consider the fields to be the real and imaginary parts of a complex field $\Phi$. We will show that it is possible to construct a model in which $|\Phi|$ is sub-Planckian throughout and $n_{s}$ and $r$ are in good agreement with Planck and BICEP2. In order to satisfy the Lyth bound, the minimum of the potential in field space cannot be in the radial direction, as it must describe a sufficiently long path across the two-field landscape. (A similar observation was made in \cite{antusch}.) The path must be aperiodic in the phase of $\Phi$ in order to be long enough to satisfy the Lyth bound.  As an existence proof of a successful two-field inflation model, we will present a simple example in which the path is nearly circular but aperiodic, so that the minimum of the potential has a spiral-like path in the complex plane.

\section{Inflation with an Aperiodic Modulated Potential}

    We will consider an potential of the form
\be{e1} V(\phi, \theta) = f_{A}(\phi, \theta) V(|\Phi|) ~,\ee
where $\Phi = (\phi/\sqrt{2})e^{i \theta}$. $f_{A}(\phi,\theta)$ is a modulating function which creates an aperiodic local minimum of the potential, such that the minimum as a function of $\theta$ describes a spiral-like path. In the following we will consider a sinusoidal modulating function, 
\be{e2} f_{A}(\phi, \theta) = 1 + \frac{A}{2} \sin
\left(\frac{|\Phi|^{m}}{\Lambda^{m}} + \theta\right)    ~. \ee
The base potential $V(|\Phi|)$ is assumed to be a $|\Phi|^{n}$ potential. Therefore  
\be{e3} V(\phi, \theta) = \frac{\lambda |\Phi|^{n}}{M_{Pl}^{n-4}} \left(1 + \frac{A}{2} \sin
\left(\frac{|\Phi|^{m}}{\Lambda^{m}} + \theta\right) \right)  ~,\ee
where $\lambda$ a free parameter and we have set the dimensional mass to $M_{Pl}$. In Figure 1 we show the potential\footnote{ In \cite{dantes}, a model ("Dante's Inferno")  based on string axion monodromy is proposed which is dynamically similar to the model discussed here. A second axion field $\theta$ is introduced, while the potential for the monodromy axion field $r$ is generalized to a $r^{p}$ potential. The dynamics of this model are essentially identical to the present model with $m = 1$. The interpretation of the fields is different, with two axion fields and a potential from string dynamics in the case of \cite{dantes} and a single complex field with the phase and modulus serving as the dynamical fields in the present model. In \cite{dantes} the modulation is due to an additive axion term in the potential,  whereas here we consider a multiplicative modulation of the potential.} as a function of $\phi$ at a fixed value of $\theta$.  

  The potential has a local minimum at $\phi(\theta)$, where 
\be{e3a} \left(\frac{\phi(\theta)}{\sqrt{2} \Lambda}\right)^{m} \approx \; 2 n \pi -\frac{\pi}{2} - \theta   ~\ee
and $n$ is an integer. This is a good approximation if $A(\phi/\sqrt{2} \Lambda)^m \gg 1$, assuming that $A$ is small compared to 1. In Figure 2 we show the spiral-like minimum in the complex plane as a function of $\theta$.  
The distance $a$ along the minimum in field space is related to $\theta$ by 
\be{e4}  da = \sqrt{\phi^{2} + \left(\frac{d\phi}{d\theta}\right)^{2}  } d \theta  ~,\ee 
From \eq{e3a}, 
\be{e5} \frac{d \phi}{d \theta} = -\left( \frac{\sqrt{2} \Lambda}{\phi} \right)^{m} \frac{\phi}{m}   ~.\ee
Therefore if $(\sqrt{2}\Lambda/\phi)^{2m} \ll 1$ then to a good approximation $da = \phi(\theta) d\theta$. In this case we can consider $a$ to be a canonically normalized field along the minimum of the potential. The model will behave as a single field inflation model if field orthogonal to $a$, $\phi$, has a mass much larger than $H$. We will show later that this can easily be satisfied.

 We can construct an effective single-field potential for the slowly-rolling field. Using $da = \phi(\theta) d\theta$, where $\phi(\theta)$ is related to $\theta$ via \eq{e3a}, we find 
\be{xx1}  \int da = \int \phi d \theta = \int \phi \left( \frac{d \theta}{d \phi} \right) d \phi = \int -m \left(\frac{\phi}{\sqrt{2} \Lambda} \right)^{m} d \phi   ~.\ee
Therefore
\be{xx2}  a = \frac{m}{\left(m+1\right)} \frac{1}{\left(\sqrt{2} \Lambda\right)^{m} } \left(\phi_{0}^{m+1} - \phi^{m+1} \right)  ~,\ee
where we have defined $a = 0$ when $\phi = \phi_{0}$. Thus
\be{xx3}  \phi = 
\left( \left(\sqrt{2} \Lambda\right)^{m} \left(\frac{m+1}{m}\right) \right)^{\frac{1}{m+1}} \left( \left(\frac{m}{m+1}\right) \frac{\phi_{0}^{m+1}}{\left(\sqrt{2} \Lambda\right)^{m} } - a\right)^{\frac{1}{m+1}}    ~.\ee
We can then define a new slow-roll field $\hat{a}$, where
\be{xx4}  \hat{a} =  \frac{m}{\left(m+1\right)} \frac{\phi_{0}^{m+1}}{\left(\sqrt{2} \Lambda\right)^{m} } - a   ~.\ee
Therefore
\be{xx4a}  \hat{a} = \frac{m}{\left(m+1\right)} \frac{\phi^{m+1}}{\left(\sqrt{2} \Lambda\right)^{m} }  ~.\ee 
Along the minimum, 
\be{xx5} V(\hat{a}) = \left(1-\frac{A}{2}\right)
\frac{\lambda \phi^{n}(\hat{a})}{ \left(\sqrt{2}\right)^{n} M_{Pl}^{n-4}} \approx \frac{\lambda \phi^{n}(\hat{a})}{ \left(\sqrt{2}\right)^{n} M_{Pl}^{n-4}} = \frac{\lambda}{\left(\sqrt{2}\right)^{n} M_{Pl}^{n-4} } 
\left( \left(\sqrt{2} \Lambda\right)^{m} \left(\frac{m+1}{m}\right) \right)^{\frac{n}{m+1}} \hat{a}^{\frac{n}{m+1}}    ~.\ee

\begin{figure}[htbp]
\begin{center}
\epsfig{file=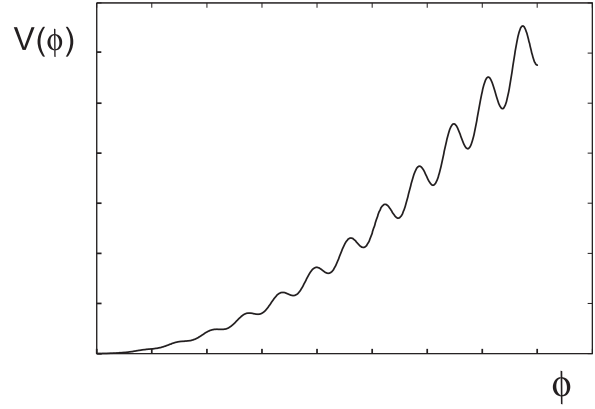, width=0.3\textwidth, angle = -90}
\caption{The modulated potential as a function of $\phi$ along fixed $\theta$.}
\label{fig1}
\end{center}
\end{figure}

\begin{figure}[htbp]
\begin{center}
\epsfig{file=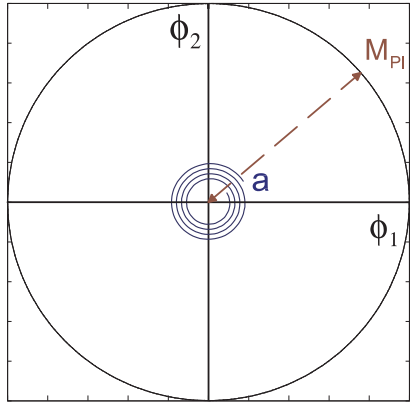, width=0.3\textwidth, angle = 0}
\caption{Aperiodic spiralling minimum of the potential. Here $\Phi = (\phi_{1} + i \phi_{2})/\sqrt{2}$. The field can be well within the Planck scale throughout while travelling a super-Planckian distance $\Delta a$ in field space. }
\label{fig2}
\end{center}
\end{figure}

    Therefore we can use the standard results for chaotic inflation models with power $n/(m+1)$. The slow-roll parameters as a function of the number of e-foldings $N$ are then 
\be{e18} \epsilon = \left(\frac{n}{4\left(m+1\right)}\right) \frac{1}{N}   ~\ee
and 
\be{e19} \eta = \left(\frac{n-m-1}{2\left(m+1\right)} \right)\frac{1}{N}   ~.\ee
The spectral index $n_{s}$ and the tensor-to-scalar ratio $r$ are 
\be{e20} n_{s} = 1 + 2 \eta - 6 \epsilon =  1 - \left(\frac{n+2m+2}{2\left(m+1\right)} \right) \frac{1}{N}    ~\ee 
and
\be{e21} r = 16 \epsilon = \left(\frac{4n}{m+1}\right) \frac{1}{N}  ~.\ee
$\hat{a}$ and $\phi$ are related to $N$ by 
\be{t1}  \frac{\hat{a}}{M_{Pl}} = \left(\frac{2nN}{m+1}\right)^{1/2}  ~\ee
and 
\be{t2} \left(\frac{\phi}{\sqrt{2} \Lambda}\right)^{m+1}
= \left(\frac{M_{Pl}^2}{\Lambda^2} \frac{n\left(m+1\right)N}{m^2} \right)^{1/2}    ~.\ee

   This model can therefore provide a good fit to the spectral index observed by Planck ($n_{s} = 0.9607 \pm 0.007$) \cite{planck} and the tensor-to-scalar ratio observed by BICEP2 ($r = 0.2^{+0.07}_{-0.05}$) \cite{bicep1,bicep2} by choosing the parameters to 
reproduce the results of $\phi^2$ chaotic inflation. In that case $n_{s} = 1-2/N$ and $r = 8/N$, which gives $n_{s} = 0.964$ and $r = 0.15$ for $N = 55$. The spectral index of our model will equal that of $\phi^2$ chaotic inflation if
\be{e22} n = 2m+2   ~.\ee
For example, if $m = 1$ then $n = 4$. 

   We next show that $|\Phi|$ is sub-Planckian throughout. To do this we need to fix the value of $\Lambda$. $\Lambda$ is determined by the adiabatic power spectrum, 
\be{e23} {\cal P}_{\zeta} = \frac{V}{24 \pi^2 \epsilon M_{Pl}^{4}}    ~.\ee
Therefore 
\be{e24} {\cal P}_{\zeta} = \frac{\lambda}{6 \pi^{2}} 
\left(\frac{m+1}{n}\right) \left(\frac{n\left(m+1\right)}{m^{2}}\right)^{\frac{n}{2\left(m+1\right)} }  \left(\frac{\Lambda}{M_{Pl}} \right)^{\frac{nm}{m+1}} N^{\frac{2m +n +2}{2\left(m+1\right)} }   ~.\ee
For the case $n = 4$ and $m = 1$, this implies that
\be{e26} \frac{\Lambda}{M_{Pl}} = \left( \frac{3 \pi^2 {\cal P}_{\zeta}}{2 \lambda} \right)^{1/2} \frac{1}{N}  ~.\ee
With $N = 55$ and ${\cal P}_{\zeta}^{1/2} = 4.8 \times 10^{-5}$, this gives
\be{e27} \frac{\Lambda}{M_{Pl}} = 3.3 \times 10^{-6} \lambda^{-1/2}     ~.\ee
From \eq{t2} we find,  
\be{e28} \frac{\phi}{\Lambda} = \left(\frac{32 N M_{Pl}^{2}}{\Lambda^{2}} \right)^{1/4}   ~\ee
therefore 
\be{e29}  \frac{\phi}{M_{Pl}} = \left(32 N\right)^{1/4} \left( \frac{\Lambda}{M_{Pl}} \right)^{1/2}   ~.\ee
At $N = 55$ this gives,
\be{e30}  \frac{\phi}{M_{Pl}} = 0.012\; \lambda^{-1/4}    ~.\ee
  Thus inflation will be sub-Planckian as long as $\lambda$ is not very small compared to 1. The natural range of values for $\lambda$ in particle physics models is $\lambda = 0.001 - 1$, in which case $|\Phi|/M_{Pl} = 0.01-0.05$.

     The Lyth bound requires that the change in the inflaton field $\Delta \hat{a}$ satisfies \cite{lythb,antusch}
\be{e36} \frac{\Delta \hat{a}}{M_{Pl}} \geq \sqrt{2 \epsilon_{min}} \; N  \approx \sqrt{\frac{r}{8}} \; N ~\ee
where $\epsilon_{min}$ is the smallest value during inflation
and $r \approx 16 \epsilon_{min}$ is assumed. With $r = 0.2 $ and $N = 55$ this requires that $\Delta \hat{a} \gae 8.7 M_{Pl}$. 
In our model $\Delta \hat{a}$ is essentially equal to $\hat{a}$ in \eq{t1}.
For $n = 4$ and $m = 1$, the change during 55 e-foldings is $\Delta \hat{a} \approx 15 M_{Pl}$, as usual for a $\phi^2$ chaotic inflation model.

We finally check the assumptions which underlie our analysis. In order to treat inflation as an effective single field model with inflaton $\hat{a}$, we require that $m_{\phi}^{2} \gg H^2$. At the local minimum the mass squared of $\phi$ is given by 
\be{e6}  m_{\phi}^2 = \frac{A m^{2} \lambda |\Phi|^{n+2m-2}}{4 M_{Pl}^{n-4} \Lambda^{2m}}   ~.\ee
Using this, the condition $m_{\phi}^{2} \gg H^2$ becomes 
\be{e32}  \left(\frac{\phi}{\sqrt{2} \Lambda}\right)^{2m-2}  \gg   \frac{4}{3 m^2 A}\frac{\Lambda^2}{M_{Pl}^2}   ~.\ee
For $m = 1$ this becomes 
\be{e33} A \gg \frac{4}{3} \left(
\frac{\Lambda}{M_{Pl}}\right)^{2}   ~.\ee
With $\Lambda/M_{Pl} = 3.3 \times 10^{-6} \lambda^{-1/2}$ this requires that $A \gg 1.5 \times 10^{-11} \lambda^{-1}$. This is easily satisfied so long as $A$ is not extremely small.

A second assumption is that $(\phi/\sqrt{2}\Lambda)^{2m} \gg 1$, in order that $da = \phi d\theta$ is true and $a$ can be treated as a canonically normalized scalar field. For $m = 1$ and $N = 55$ we find that $\phi/\sqrt{2} \Lambda \approx 2500 \lambda^{1/4}$. This is much larger than 1 so long as $\lambda$ is not extremely small. 

Finally, we require that $A (\phi/\sqrt{2} \Lambda)^{m} \gg 1$ in order that \eq{e3a} is valid. For $m = 1$ this will be satisfied if $A(2500 \;\lambda^{1/4}) \gg 1$, which will be true if $A \; \lambda^{1/4}$ is not too small compared to 1.

 In summary, the model can generate inflation which is in good agreement with observations while keeping $|\Phi|$ sub-Planckian throughout. For the case $n = 4$ and $m = 1$ the  potential is 
\be{e31} V(\phi, \theta) = \lambda |\Phi|^{4} \left(1 + \frac{A}{2} \sin
\left(\frac{|\Phi|}{\Lambda} + \theta\right) \right)  ~,\ee
where $\Phi = \phi e^{i \theta}/\sqrt{2}$, and the results for $n_{s}$ and $r$ are identical to the case of $\phi^2$ chaotic inflation. With $\lambda = 1$, in which case 
$\Lambda = 3.3 \times 10^{-6} M_{Pl} = 7.9 \times 10^{12} \GeV$, 
the value of $|\Phi|$ satisfies $|\Phi| \lae 0.01 M_{Pl}$ throughout the final $N = 55$ e-foldings. The results are independent of the value of $A$ so long as $A$ is small compared to 1. 

  A $|\Phi|^4$ potential is a natural choice for the base potential from a dimensional point of view. Assuming that $m$ is an integer, the simplest models then correspond to $m =1$ and $m = 2$, which behave as $\phi^2$ and $\phi^{4/3}$ chaotic inflation. The former is good agreement with present BICEP2 results but in tension with the Planck upper bound, $r < 0.11$ (95$\%$ c.l.) \cite{planck}. The $m = 2$ model predicts $n_{s} = 1-5/3N = 0.970 $ and $r = 16/3N = 0.097$ when $N = 55$. (In this case $|\Phi|/M_{Pl} \lae 0.008 \lambda^{-1/4}$ during inflation.) Therefore if the forthcoming Planck result for $r$ turns out to be substantially less than the BICEP2 value, then this model could become favoured.

\section{Conclusions}

     In the absence of a shift or other symmetry, the Lyth bound presents a severe problem for single-field inflation if Planck-suppressed corrections to the potential exist. N-flation would require a very large number of scalar fields to keep the inflaton sub-Planckian. The alternative is to consider multi-field inflation with a trajectory which is non-radial, so that the {\it change}  in the inflaton field can be super-Planckian while the magnitude of the field remains sub-Planckian throughout. This requires an appropriate potential landscape. We have presented an explicit example of such a model. The model is based on a complex field with a nearly circular but aperiodic trajectory in the complex plane. This is achieved by a sinusoidal modulation of a $|\Phi|^{n}$ potential. The model acts as a single-field inflation model with a spiral-like trajectory. We have shown that the simplest sinusoidal modulation of a $|\Phi|^4$ potential exactly reproduces the values for $n_{s}$ and $r$ of the $\phi^2$ chaotic inflation model while maintaining $|\Phi| \lae 0.01 M_{Pl}$ throughout.  More generally, the simplest models based on a $|\Phi|^4$ potential favour predictions equivalent to $\phi^2$ and $\phi^{4/3}$ chaotic inflation. 
  
   The model therefore serves as an existence proof for two-field inflation models that are consistent with Planck and BICEP2 and which satisfy the Lyth bound while remaining sub-Planckian throughout. The aperiodic modulated $|\Phi|^4$ potential we have studied has no obvious physical explanation at present. On the other hand, the potential is a particularly simple example of a two-field potential landscape and might therefore be realized as part of a complete theory of Planck-scale physics.   

    We finally comment on the relation of this model to axion inflation and monodromy \cite{dantes,axmon1,axmon2,westp,setax1,setax2,ido1,ido2}. In the axion inflation monodromy model of \cite{axmon1} and \cite{axmon2}, the effective theory consists of two real scalar fields with a broken shift symmetry. (The dynamics of monodromy in the later string versions are not explicit in the four-dimensional effective theory \cite{westp}.) The decay constants of the axions are approximately aligned to produce an almost flat direction which can be used for inflation.   
The effective theory is equivalent to a single-field inflation model with a super-Planckian inflaton and Planck-suppressed corrections eliminated by the shift symmetry. The sub-Planckian nature of the model is only apparent in the full theory with two complex fields, where the two real scalar fields are seen to be two axions and the full theory is therefore sub-Planckian if their decay constants are sufficiently sub-Planckian. In our model, the sub-Planckian nature of the model is explicit in the effective theory itself; in the simplest $n = 4$, $m = 1$ model the real scalar fields $\phi_{1}$ and $\phi_{2}$ are no larger than O(0.01)$M_{Pl}$. Our model also differs in being based on a single complex field, with the modulus of the field playing a role in the aperiodic potential. The axion inflation models of \cite{axmon1} and \cite{axmon2} make predictions which are generally different from the power-law chaotic inflation predictions of our model. The string axion monodromy models of \cite{westp,setax1,setax2}, on the other hand,  are equivalent to chaotic inflation models with $V(\phi) \propto \phi$ and an additional sinusoidal modulation. (Recently this has been generalized to a range of power-law potentials \cite{west2}.) Finally, the two-axion "Dante's Inferno" model \cite{dantes}   
makes predictions which are equivalent to those of the present model for the case $m = 1$.

\section*{Acknowledgements} Thanks to David Lyth for his comments.  The work of JM is partially supported by STFC grant
ST/J000418/1.

\renewcommand{\theequation}{A-\arabic{equation}}
 \setcounter{equation}{0} 

\end{document}